\documentclass[journal]{IEEEtran}
\usepackage{multirow}
\usepackage{multicol}
\usepackage{lipsum}
\usepackage{graphicx}
\usepackage{graphics}
\usepackage{amsmath}
\usepackage{amsbsy}
\usepackage{blindtext}
\usepackage{tcolorbox}
\usepackage{amssymb}
\usepackage{scalerel}
\usepackage{mathabx}
\usepackage{verbatim}
\usepackage{booktabs}
\usepackage{rotating,tabularx}
\usepackage{mathrsfs} 

\usepackage[percent]{overpic} 
\usepackage{xcolor}

\usepackage{hyperref}

\usepackage{enumitem}

\usepackage{caption}
\usepackage{adjustbox}
\usepackage{soul}
\usepackage{xcolor}
\usepackage{cite}
\usepackage{tikz}
\usepackage{color, colortbl}
\usepackage[first=0,last=9]{lcg}
\definecolor{LightGray}{rgb}{0.7,0.7,0.7}
\usepackage{pifont}

\usepackage[wby]{callouts}
\usetikzlibrary{shapes.multipart}

\usepackage{array}
\usepackage{makecell}

\allowdisplaybreaks

\usepackage{amsthm}
\theoremstyle{definition}

\theoremstyle{remark}

\usepackage[utf8]{inputenc}
\usepackage[english]{babel}

\usepackage[caption=false,font=footnotesize]{subfig}
\usepackage[T1]{fontenc}
\usepackage{scalerel,stackengine}
\newcommand\reallywidecheck[1]{%
\savestack{\tmpbox}{\stretchto{%
  \scaleto{%
    \scalerel*[\widthof{\ensuremath{#1}}]{\kern-.6pt\bigwedge\kern-.6pt}%
    {\rule[-\textheight/2]{1ex}{\textheight}}
  }{\textheight}%
}{0.5ex}}%
\stackon[1pt]{#1}{\scalebox{-1}{\tmpbox}}%
}
\stackMath
\IEEEoverridecommandlockouts
\IEEEoverridecommandlockouts

\renewcommand{\baselinestretch}{1}

\newcommand{\cmark}{\ding{51}}  
\newcommand{\xmark}{\ding{55}}  
\usepackage{array}        
\usepackage{booktabs}     
\usepackage{multirow}     
\usepackage{amssymb}      
\usepackage{pifont}       

\newcommand*{\rn}{\textcolor{black}}

\newcommand*{\dkm}{\textcolor{black}}

\newif\ifarxiv
\arxivtrue

\begin{document}

\title{\LARGE\bf
Certifying the Nonexistence of Feasible Paths\\ Between Power System Operating Points
}

\author{Mohammad Rasoul Narimani,$^{\ast}$ 
Katherine R. Davis,$^{**}$ and Daniel K. Molzahn$^{\dagger}$
\thanks{${\ast}$: Electrical and Computer Eng., California State Univ. Northridge. \mbox{rasoul.narimani@csun.edu}. Support from NSF \#2308498 and \#2523881.}%
\thanks{${**}$: Electrical and Computer Eng., Texas A\&M Univ. katedavis@tamu.edu.}
\thanks{$^{\dagger}$: Electrical and Computer Eng., Georgia Institute of Technology. molzahn@gatech.edu. Support from NSF \#2023140.}%
}

\maketitle

\begin{abstract}
By providing the optimal operating point that satisfies both the power flow equations and engineering limits, the optimal power flow (OPF) problem is central to power systems operations. While extensive research has focused on computing high-quality OPF solutions, assessing the feasibility of transitioning between operating points remains challenging since the feasible spaces of OPF problems may consist of multiple disconnected components. It is not possible to transition between operating points in different disconnected components without violating OPF constraints. To identify such situations, this paper introduces an algorithm for certifying the infeasibility of transitioning between two operating points within an OPF feasible space. As an indication of potential disconnectedness, the algorithm first seeks an infeasible point on the line connecting a pair of feasible points. The algorithm then certifies disconnectedness by using convex relaxation and bound tightening techniques to show that all points on the plane that is normal to this line are infeasible. Using this algorithm, we provide the first certifications of disconnected feasible spaces for a variety of OPF test cases. 

\end{abstract}

\section{Introduction}
\label{Introduction}


Optimal power flow (OPF) is a crucial problem in power system operations. This problem seeks an optimal operating point 
within a feasible space determined by the power flow equations and inequality constraints that limit voltage magnitudes, line flows, and generator outputs.
OPF feasible spaces are nonconvex and may consist of multiple disconnected components~\cite{molzahn_opf_spaces}, making the OPF problem NP-hard in general~\cite{bienstock2015nphard}.

A wide variety of local optimization and approximation algorithms have been applied to OPF problems~\cite{opf_litreview1993IandII,molzahn2017survey}. 
Many convex relaxations have also been developed to bound optimal objective values, certify infeasibility, and in some cases, obtain globally optima~\cite{molzahn2017survey}. An OPF problem's difficulty is closely related to the convexity characteristics of the feasible space. In this context, many research efforts have studied the geometry of the OPF feasible spaces, e.g.,~\cite{bernie_opfconvexity,zhang2011,bukhsh_tps,Narimani_ACC,molzahn_opf_spaces,molzahn2017survey,molzahn-nonconvexity_search,chiang2018}. 

The growth in fluctuating renewable generation motivates the need to manage increasingly frequent transitions between operating points to avoid large constraint violations (e.g., voltages and flows exceeding limits).
There is limited literature on this topic. Prior work focuses on computing feasible paths between operating points within non-convex OPF feasible spaces~\cite{lee2020,barros2022,turizo2025}. Using power flow linearizations, related work in~\cite{F_Capitanescu, D_T_Phan} seeks the fewest control actions necessary to actuate a transition between operating points. However, the failure of existing algorithms to identify a feasible path does not guarantee that no path exists, i.e., they do not ensure that the initial and target operating points are necessarily in different disconnected components of the OPF feasible space.

In this context, we propose an algorithm which certifies that the feasible space of an OPF problem is disconnected. This algorithm contributes to the literature on the characteristics of OPF feasible spaces (e.g.,~\cite{bernie_opfconvexity,zhang2011,bukhsh_tps,Narimani_ACC,molzahn_opf_spaces,molzahn2017survey,molzahn-nonconvexity_search,chiang2018}) and identifies when feasible path algorithms such as~\cite{lee2020,barros2022,turizo2025} will necessarily fail. \dkm{This avoids the computational effort of running path finding algorithms in situations where no feasible path exists. Moreover, certifying the disconnectedness of an OPF feasible space can inform operators of situations where constraint violations are unavoidable during the transition to a target operating point. As large constraint violations can damage equipment or trigger protection systems, this would help operators know when to use, e.g., decomposition and convex restriction methods~\cite{D_Lee} to identify a target operating point that is within the same component of the feasible space as the initial point.}

Our proposed algorithm leverages concepts from~\cite{molzahn-nonconvexity_search} which identifies and characterizes non-convexities within OPF feasible spaces. The algorithm in~\cite{molzahn-nonconvexity_search} searches for a pair of feasible points connected by a line segment containing an infeasible point. The existence of such points certifies the presence of a non-convex region within an OPF problem's feasible space, but does not necessarily indicate that the feasible space is disconnected. Note that the causes of non-convexities in OPF feasible spaces are empirically investigated in~\cite{Narimani_ACC}, with results showing that non-convexities are often associated with binding lower bounds on voltage magnitudes and reactive generation. 


The algorithm proposed in this paper first uses the approach in~\cite{molzahn-nonconvexity_search} to identify two candidate operating points which may be in different disconnected components due to the presence of an infeasible point between them. Our algorithm then formulates an optimization problem which augments the OPF constraints with a constraint restricting its solution to a plane that is perpendicular to the line connecting the two feasible points and passes through the previously identified infeasible point. Infeasibility of this optimization problem certifies that the OPF problem's feasible space is disconnected. To prove infeasibility, we relax this optimization problem using the Quadratic Convex (QC) power flow relaxation~\cite{coffrin2015qc} and apply optimization-based bound tightening techniques~\cite{coffrin2016strengthen_tps,StrongSOCPRelaxations,chen2015,arctan2}. To the best of our knowledge, this is the first algorithm that can rigorously identify the presence of disconnected components in OPF feasible spaces. Our numerical results show that several challenging OPF test cases have disconnected feasible spaces.


This paper is organized as follows. Section~\ref{OPF_overview} reviews the OPF formulation. Section~\ref{sec:QC relaxation overview} describes the QC relaxation and optimization-based bound tightening. Section~\ref{sec:proposed_algorithm} presents the proposed algorithm for determining the disconnectedness of OPF feasible spaces. Section~\ref{Numerical_results} empirically demonstrates this algorithm. Section~\ref{conclusion} concludes the paper. 
\ifarxiv
\else
An extended version in~\cite{narimani2025certifyingnonexistencefeasiblepath} details numeric values for the test cases.
\fi
\section{Optimal Power Flow Overview}
\label{OPF_overview}
This section overviews the AC~OPF problem. Consider an $n$-bus system, where $\mathcal{N} = \left\lbrace 1, \ldots, n \right\rbrace$, $\mathcal{G}$, and $\mathcal{L}$ are the sets of buses, generators, and lines. Let ${P}_{d,i} + \mathbf{j}{Q}_{d,i}$ and ${P}_{g,i} + \mathbf{j} {Q}_{g,i}$ represent the active and reactive load demand and generation, respectively, at bus~$i\in\mathcal{N}$, where $\mathbf{j} = \sqrt{-1}$. Let $g_{sh,i} + \mathbf{j} b_{sh,i}$ denote the shunt admittance at bus $i$. Let $V_i$ and $\theta_i$ represent the voltage magnitude and angle at bus~$i\in\mathcal{N}$. For each generator $i\in\mathcal{G}$, define a quadratic generation cost function with coefficients $c_{2,i} \geq 0$, $c_{1,i}$, and $c_{0,i}$. Denote $\theta_{lm}=\theta_{l}-\theta_{m}$. Specified upper and lower limits are denoted by $\left(\overline{\,\cdot\,}\right)$ and $\left(\underline{\,\cdot\,}\right)$, respectively. Buses $i\in\mathcal{N}\setminus\mathcal{G}$ have generation limits set to zero. 

Each line $\left(l,m\right)\in\mathcal{L}$ is modeled as a $\Pi$ circuit with mutual admittance $g_{lm}+\mathbf{j} b_{lm}$ and shunt admittance $\mathbf{j} b_{sh,lm}$. (Our approach is applicable to more general line models, such the M{\sc atpower}~\cite{matpower} model that allows for off-nominal tap ratios and non-zero phase shifts.) Let $p_{lm}$, $q_{lm}$, and $\overline{s}_{lm}$ represent the active and reactive power flows and the maximum apparent power flow limit on the line that connects buses~$l$ and~$m$.

Using these definitions, the OPF problem is
\begin{subequations}
\label{OPF formulation}
\begin{align}
\label{eq:objective}
& \min\quad \sum_{{i}\in \mathcal{G}}
c_{2i}\left(P_{g,i}\right)^2+c_{1i}\,P_{g,i}+c_{0i}\\ 
&\nonumber \text{subject to} \quad \left(\forall i\in\mathcal{N},\; \forall \left(l,m\right) \in\mathcal{L}\right) \\
\label{eq:pf1}
& P_{g,i}-P_{d,i} = g_{sh,i}\, V_i^2+\sum_{\substack{(l,m)\in \mathcal{L}\\ \text{s.t.} \hspace{3pt} l=i}} p_{lm}+\sum_{\substack{(l,m)\in \mathcal{L}\\ \text{s.t.} \hspace{3pt} m=i}} p_{ml}, \\
\label{eq:pf2}
& Q_{g,i}-Q_{d,i} = -b_{sh,i}\, V_i^2+\sum_{\substack{(l,m)\in \mathcal{L}\\ \text{s.t.} \hspace{3pt} l=i}} q_{lm}+\sum_{\substack{(l,m)\in \mathcal{L}\\ \text{s.t.} \hspace{3pt} m=i}} q_{ml},\\[-10pt]
\label{eq:OPF1}
& \theta_{ref}=0,\\
\label{eq:OPF2}
& \underline{P}_{g,i}\leq P_{g,i}\leq \overline{P}_{g,i}, ~~~\underline{Q}_{g,i}\leq Q_{g,i}\leq \overline{Q}_{g,i},\\
\label{eq:OPF4}
& \underline{V}_i\leq V_{i} \leq \overline{V}_i,~~~\underline{\theta}_{lm}\leq \theta_{lm}\leq \overline{\theta}_{lm},\\
\label{eq:pik}
&p_{lm} = g_{lm} v_l^2 - g_{lm} V_l V_m\cos\left(\theta_{lm}\right) - b_{lm} V_l V_m\sin\left(\theta_{lm}\right), \\[-8pt]
\label{eq:qik}
& \nonumber q_{lm} = -\left(b_{lm}+b_{sh,lm}/2\right) V_l^2 + b_{lm} V_l V_m\cos\left(\theta_{lm}\right)\\ &\qquad\qquad  - g_{lm} V_l V_m\sin\left(\theta_{lm}\right), \\
\label{eq:OPF8}
& \left(p_{lm}\right)^2+\left(q_{lm}\right)^2 \leq \left(\overline{s}_{lm}\right)^2,  \left(p_{ml}\right)^2+\left(q_{ml}\right)^2 \leq \left(\overline{s}_{lm}\right)^2.
\end{align}
\end{subequations}
The objective~\eqref{eq:objective} minimizes the generation cost. Constraints~\eqref{eq:pf1} and~\eqref{eq:pf2} enforce power balance at each bus. Constraint~\eqref{eq:OPF1} sets the angle reference. Constraints~\eqref{eq:OPF2}--\eqref{eq:OPF4} limit the power generation, voltage magnitudes, and angle differences. Constraints~\eqref{eq:pik}--\eqref{eq:qik} relate the voltages and power flows on each line, and \eqref{eq:OPF8} limits apparent power flows.

\section{Review of the QC Relaxation and Optimization-based Bound Tightening}
\label{sec:QC relaxation overview}
We next review the QC relaxation~\cite{coffrin2015qc} and optimization-based bound tightening~\cite{coffrin2016strengthen_tps,StrongSOCPRelaxations,chen2015,arctan2} that underlie our algorithm.

\subsection{Formulation of the QC Relaxation}
\label{sec:QC_relaxation}
The QC relaxation convexifies the OPF problem~\eqref{OPF formulation} by enclosing non-convex terms 
in convex envelopes. This relaxation defines new variables $w_{ii}$, $w_{lm}$, $c_{lm}$, and $s_{lm}$ for products of voltage magnitudes and products of voltage magnitudes and trigonometric terms, i.e.,   
$V_lV_m\cos(\theta_{lm})$ and $V_lV_m\sin(\theta_{lm})$:
%
\label{eq:cs}
\begin{align}
\nonumber & w_{ii} = V_i^2,\; \forall i \in\mathcal{N}, \quad w_{lm} = V_l V_m, \quad c_{lm} =  w_{lm} \cos\left(\theta_{lm} \right), \\
& \qquad s_{lm}  = w_{lm}\sin\left(\theta_{lm} \right),\;   \forall \left(l,m\right) \in\mathcal{L}.
\end{align}
For every $\left(l,m\right)\in\mathcal{L}$, these definitions imply the following relationships among the variables $w_{ll}$, $c_{lm}$, and $s_{lm}$:
\begin{subequations}
\label{eq:cs_relationships}
\begin{align}
\label{eq:Jabr}
& c_{lm}^2+s_{lm}^2=w_{ll}w_{mm},\\
\label{eq:cs_relationships_cs}
& c_{lm}=c_{ml}, \quad s_{lm}=-s_{ml}.
\end{align}
\end{subequations}
Following~\cite{Jabr2006},~\eqref{eq:Jabr} is relaxed to a convex second-order cone constraint by replacing the equality with an inequality:
\begin{equation}
\label{eq:Jabr_relaxed}
c_{lm}^2+s_{lm}^2 \leq w_{ll}w_{mm}.
\end{equation}
The QC relaxation encloses the squared and bilinear product terms in convex envelopes, shown here as set-valued functions:%
\begin{subequations}
\label{eq:product_envelopes}
\begin{align}
\label{eq:squareenvelopes}
\langle x^2\rangle^T =
\begin{cases}
\widecheck{x}: \begin{cases}\check{x} \geq x^2,\\
\widecheck{x} \leq \left({\overline{x}+\underline{x}}\right) x-{\overline{x} \underline{x}}.\\
\end{cases}
\end{cases}\\
\label{eq:mccormick}
\langle {xy}\rangle^M  =
\begin{cases}
\widecheck{xy}:\begin{cases}
\widecheck{xy} \geq {\underline{x}} y+ {\underline{y}} x-{\underline{x} \underline{y}},\\
\widecheck{xy} \geq {\overline{x}} y+ {\overline{y}} x-{\overline{x} \overline{y}},\\
\widecheck{xy} \leq {\underline{x}} y+ {\overline{y}} x-{\underline{x}} {\overline{y}},\\
\widecheck{xy} \leq {\overline{x}} y+ {\underline{y}} x-{\overline{x} \underline{y}}.\\
\end{cases}
\end{cases}
\end{align}
\end{subequations}
Here, $\widecheck{x}$ and $\widecheck{xy}$ stand for ``dummy'' variables symbolizing their respective sets. The envelope $\langle x^2\rangle^T$ is the convex hull of the square function. The McCormick envelope denoted as $\langle {xy}\rangle^M$ is the convex hull of a bilinear product~\cite{mccormick1976}.

The QC relaxation additionally uses convex envelopes $\left\langle\sin\left(x\right) \right\rangle^S$ and $\left\langle\cos\left(x\right) \right\rangle^C$ for the trigonometric functions:
\begin{subequations}
\label{eq:convex_envelopes_sin&cos}
\begin{align}
\label{eq:sine envelope}
\nonumber &\left\langle \sin(x)\right\rangle^S =\\
&\; \begin{cases}
\widecheck{S}:\begin{cases}
\widecheck{S}\leq\cos\left(\frac{x^m}{2}\right)\left(x-\frac{x^m}{2}\right)+\sin \left(\frac{x^m}{2}\right),\\
\widecheck{S}\geq\cos\left(\frac{x^m}{2}\right)\left(x+\frac{x^m}{2}\right)-\sin\left(\frac{x^m}{2}\right),\\
\widecheck{S}\geq\frac{\sin\left({\underline{x}}\right)-\sin\left(\overline{x}\right)}{{\underline{x}-\overline{x}}}\left(x-{\underline{x}}\right)+\sin\left({\underline{x}}\right) \text{if~} \underline{x}\geq0,\\
\widecheck{S}\leq\frac{\sin\left({\underline{x}}\right)-\sin\left({\overline{x}}\right)}{{\underline{x}-\overline{x}}}\left(x-{\underline{x}}\right)+\sin\left({\underline{x}}\right) \text{if~} {\overline{x}}\leq0.
\end{cases}
\end{cases}\raisetag{3.5em}\\
\label{eq:cosine envelope}
&\nonumber\left\langle\cos(x)\right\rangle^C=\\
&\;\begin{cases}
\widecheck{C}:\begin{cases}
\widecheck{C}\leq 1-\frac{1-\cos\left({x^m}\right)}{\left(x^m\right)^2}x^2,\\
\widecheck{C}\geq\frac{\cos\left(\underline{x}\right)-\cos\left({\overline{x}}\right)}{{\underline{x}-\overline{x}}}\left(x-{\underline{x}}\right)+\cos\left({\underline{x}}\right).  
\end{cases}
\end{cases}
\end{align}
\end{subequations}
%
Here, $x^m= \max(\left|\underline{x}\right|,\left|\overline{x}\right|)$. The variables $\check{S}$ and $\check{C}$ are placeholders for their respective sets. When $-90^\circ < \underline{x} < \overline{x} < 90^\circ$, bounds on the sine and cosine functions are as follows:
%
\begin{subequations}
\begin{align}
& \underline{s} = \sin\left(\underline{x}\right) \leq \sin(x) \leq \overline{s} = \sin\left(\overline{x}\right), \\
\nonumber & \underline{c} = \min\left(\cos(\underline{x}),\cos(\overline{x})\right) \leq \cos(x) \\ & \quad \leq \overline{c} \!=\! \begin{cases} \max\left(\cos(\underline{x}),\cos(\overline{x})\right),\; \text{if~} \mathrm{sign}\left(\underline{x}\right) \!=\!  \mathrm{sign}\left(\overline{x}\right), \\ 1, \text{~otherwise}. \end{cases}\raisetag{1em}
\end{align}
\end{subequations} 

With a slight abuse of notation, the QC relaxation substitutes the square, product, and trigonometric terms in~\eqref{OPF formulation} with the variables $w_{ii}$, $w_{lm}$, $c_{lm}$, and $s_{lm}$ in these envelopes as follows:%
\begin{subequations}
\label{eq:qc}
\begin{align}
& \min \quad \sum_{{i}\in \mathcal{G}} c_{2i}\left(P_{g,i}\right)^2+c_{1i}\,P_{g,i}+c_{0i}\\ 
&\nonumber \text{subject to} \quad \left(\forall i\in\mathcal{N},\; \forall \left(l,m\right) \in\mathcal{L}\right) \\
\label{eq:qc_p}
& P_{g,i}-P_{d,i} = g_{sh,i}\, w_{ii}+\sum_{\substack{(l,m)\in \mathcal{L}\\ \text{s.t.} \hspace{3pt} l=i}} p_{lm}+\sum_{\substack{(l,m)\in \mathcal{L}\\ \text{s.t.} \hspace{3pt} m=i}} p_{ml}, \\
\label{eq:qc_q}
& Q_{g,i}-Q_{d,i} = -b_{sh,i}\, w_{ii}+\sum_{\substack{(l,m)\in \mathcal{L}\\ \text{s.t.} \hspace{3pt} l=i}} q_{lm}+\sum_{\substack{(l,m)\in \mathcal{L}\\ \text{s.t.} \hspace{3pt} m=i}} q_{ml},\\[-0.75em]
\label{eq:qc_V}
&  (\underline{V}_i)^2\leq w_{ii} \leq (\overline{V}_i)^2,\\
\label{eq:qc_pik}
& p_{lm} = g_{lm} w_{ll} - g_{lm} c_{lm} - b_{lm} s_{lm}, \\
\label{eq:qc_qik}
& q_{lm} = -\left(b_{lm}+b_{sh,lm}/2\right) w_{ii} + b_{lm} c_{lm}- g_{lm} s_{lm}, \\
\nonumber &  w_{ii} \in\left\langle V_i^2 \right\rangle^T, 
w_{lm} \in \left\langle V_l V_m \right\rangle^M, 
c_{lm} \in \left\langle w_{lm}\left\langle\cos\left(\theta_{lm} \right)\right\rangle^C\right\rangle^M\!\!\!\!, \\[-0.5em] 
\label{eq:qc_w_c_s} & \qquad s_{lm} \in \left\langle w_{lm} \left\langle\sin\left(\theta_{lm} \right)\right\rangle^S\right\rangle^M\!\!\!\!, \\
\label{eq:qc_others}
&  \text{Equations~}\eqref{eq:OPF1}\text{--}\eqref{eq:OPF4},\,\eqref{eq:OPF8},\,\eqref{eq:cs_relationships_cs}, \,\eqref{eq:Jabr_relaxed}.
\end{align}
\end{subequations}
Note that the product terms in~\eqref{eq:pik} and~\eqref{eq:qik} are addressed in~\eqref{eq:qc_w_c_s} 
through a recursive application of \mbox{McCormick} envelopes~\eqref{eq:mccormick}.
The optimization problem~\eqref{eq:qc} is a second-order cone program (SOCP), which is convex and can be solved efficiently using commercial tools (e.g., Gurobi, Mosek, etc.).

Convex relaxations have several complementary advantages over applying local optimization methods to the non-convex OPF problem~\eqref{OPF formulation}~\cite{molzahn2017survey}. In our context, we leverage the fact that infeasibility of the QC relaxation~\eqref{eq:qc} is sufficient to ensure infeasibility of the original non-convex OPF problem~\eqref{OPF formulation}. In fact, convex relaxations are the only approach capable of rigorously certifying infeasibility of non-convex OPF problems. 

\subsection{Bound Tightening}
\label{subsec:bt}
The QC relaxation's tightness is closely linked to the precision of the bounds on voltage magnitudes ($\underline{V}_i$, $\overline{V}_i$), and angle differences ($\underline{\theta}_{lm}$, $\overline{\theta}_{lm}$). The dataset's values for these limits are often much larger than what is actually attainable considering the limitations implicit from other constraints. In other words, some limits are never binding. Leveraging this insight, algorithms focused on enhancing bound accuracy provide tighter bounds that enhance the QC relaxation's quality~\cite{coffrin2016strengthen_tps,StrongSOCPRelaxations,chen2015,arctan2}.

We use the optimization-based bound tightening algorithm from~\cite{coffrin2016strengthen_tps}. This iterative algorithm minimizes and maximizes each squared voltage magnitude and angle difference variable while enforcing the QC relaxation's constraints. To illustrate, we next examine the upper voltage limit at bus~$1$:
\begin{align}
\label{eq:bt_example}
w_{11}^\ast = \max \quad w_{11} \quad \text{subject to}\quad \eqref{eq:qc_p}\text{--}\eqref{eq:qc_others}.
\end{align}
The value $w_{11}^\ast$ establishes an upper limit on the maximum attainable value of $\left(V_{1}\right)^2$ within the feasible space. If $w_{11}^\ast$ is less than $\left(\overline{V}_{1}\right)^2$, then~\eqref{eq:bt_example} yields a smaller value of $\sqrt{w_{11}^\ast}$, serving as a tighter upper bound for $V_1$, thereby tightening the QC relaxation. As the act of tightening a bound on any particular variable may improve the achievable bounds for other variables, the bound tightening procedure follows an iterative procedure until no further bounds can be refined. 

Improving the variable bounds tightens the feasible space of the relaxed OPF problem such that the relaxation is capable of certifying infeasibility for a broader set of OPF problems. Importantly, since the QC relaxation constraints enforced in~\eqref{eq:bt_example} admit all feasible points for~\eqref{OPF formulation}, note that the tightened bounds will never cut off portions of the original non-convex OPF problem's feasible space. Thus, infeasibility of the bound-tightened QC relaxation still ensures infeasibility of the original non-convex OPF problem~\eqref{OPF formulation}. As described in the next section, we leverage this in our algorithm for certifying the presence of disconnected components in OPF feasible spaces.



\section{An Algorithm for Certifying Disconnectedness of an OPF Feasible Space}
\label{sec:proposed_algorithm}
This section presents our algorithm for certifying the presence of disconnected components in the feasible space of OPF problems. We first summarize our prior method from~\cite{molzahn-nonconvexity_search} which identifies a pair of feasible operating points which have an infeasible point in between. We then describe our proposed algorithm for using these points to certify disconnectedness. 

\subsection{Identifying Non-Convexities in OPF Feasible Spaces}
\label{sec:Identifying_nonconvexities}
\begin{figure}[b]
    \centering
    \vspace{-1em}
{\includegraphics[scale=0.72]{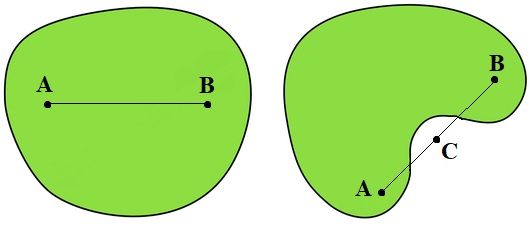}}
	\caption{Illustrations of convex and non-convex feasible regions. On the left, all points along the line segment between feasible points $A$ and $B$ are feasible. This is true for all possible pairs of feasible points. In contrast, on the right, the infeasible point $C$ is on the line between points $A$ and $B$. Points $A$, $B$, and $C$ confirm non-convexity of this region.}
\label{fig:nonconvexset}
\end{figure}

To identify two feasible operating points with an infeasible point in between, we rely on the concept of convexity. A feasible space is convex if and only if it includes all points along the line segments that connect every pair of feasible points. We utilize the algorithm from~\cite{molzahn-nonconvexity_search} to locate an infeasible point, denoted as point $C$, on the line connecting two feasible points, denoted as points $A$ and $B$, within the feasible space of the OPF problem. Fig.~\ref{fig:nonconvexset} provides illustrative examples of convexity characteristics for both convex and non-convex sets. 

In the domain of active power generation and voltage magnitudes, any point on the line segment connecting points $A$ and $B$ has active power generation at non-slack generator buses described by the expression $\lambda P_g^{A} + (1-\lambda) P_g^{B}$, where $\lambda$ is a scalar value in the range [0,1]. Similarly, the generator bus voltage magnitudes along this segment are characterized by $\lambda V^{A} + (1-\lambda) V^{B}$.
To identify non-convexity, the method in~\cite{molzahn-nonconvexity_search} seeks values for $P_g^{A}$, $P_g^{B}$, $V^{A}$, $V^{B}$, and $\lambda$ such that:
\begin{itemize}[leftmargin=*]
\item There exist power flow solutions corresponding to $P_g^{A}, V^A$ and $P_g^{B}, V^B$ that are feasible for \dkm{\emph{all} OPF constraints~\eqref{eq:pf1}--\eqref{eq:OPF8}, including voltage, reactive power, and flow limits}.

\item All power flow solutions that correspond to the point $P_g^C = \lambda P_g^{A}+ (1-\lambda) P_g^{B}$, $V^C = \lambda V^{A}+ (1-\lambda) V^{B}$ are infeasible for the OPF constraints~\eqref{eq:pf1}--\eqref{eq:OPF8}.
\end{itemize}
The method in~\cite{molzahn-nonconvexity_search} uses local optimization solvers to seek such points $A$, $B$, and $C$ for a given OPF problem.


\subsection{Proposed Algorithm for Certifying Disconnectedness}
\label{subsec:algorithm}
If the method from~\cite{molzahn-nonconvexity_search} succeeds, the resulting points $A$, $B$, and $C$ identify a non-convexity in the OPF feasible space. Since a disconnected region is necessarily non-convex, the points $A$, $B$, and $C$ may (but do not necessarily) indicate disconnectedness. To certify disconnectedness, we seek to further show that the feasible points $A$ and $B$ are separated by a hyperplane on which all points are infeasible with respect to the \dkm{OPF constraints~\eqref{eq:pf1}--\eqref{eq:OPF8}, including voltage, reactive power, and flow limits}. In that case, since all feasible paths between points $A$ and $B$ must pass through this hyperplane, no feasible path exists and the points $A$ and $B$ thus belong to different disconnected components of the OPF feasible space. 

While any separating hyperplane would suffice, we leverage knowledge regarding the infeasibility of point $C$ by selecting the hyperplane perpendicular to the line segment between points $A$ and $B$ that passes through point $C$. This ensures that at least one point on the hyperplane is infeasible. Note that our numerical results in Section~\ref{Numerical_results} also consider various rotations when the hyperplane that is perpendicular to the line between points $A$ and $B$ fails to certify disconnectedness. 

To certify infeasibility for all points on the hyperplane, we formulate a feasibility problem consisting of the OPF problem's constraints~\eqref{eq:pf1}--\eqref{eq:OPF8} augmented with an additional constraint corresponding to the hyperplane. We then apply the QC relaxation to the constraints~\eqref{eq:pf1}--\eqref{eq:OPF8} as described in Section~\ref{sec:QC_relaxation} to obtain the set of constraints
\begin{align}
\label{feasibility_problem}
\eqref{eq:qc_p}\text{--}\eqref{eq:qc_others}\text{~and~} n^\intercal \begin{bmatrix}P_g - P_g^C\\\hat{w} - (V^C)^2 \end{bmatrix} = 0,
\end{align}
where $n$ is the vector for the line between points $A$ and $B$, i.e., $n = \begin{bmatrix}P_g^A - P_g^B\\ (V^A)^2 - (V^B)^2 \end{bmatrix}$, and $\hat{w}$ denotes the vector consisting of $w_{ii}$ for all $i\in\mathcal{G}$ corresponding to the squared voltage magnitudes at each generator bus. Since the restriction to the hyperplane is a linear constraint,~\eqref{feasibility_problem} is an SOCP. \rn{While the separating hyperplane itself is defined in terms of active power generation and squared voltage magnitudes, the proposed algorithm enforces \emph{all} OPF constraints, including those involving load-bus voltages, phase angles, reactive power injections, and line flows. In other words, the feasibility problem~\eqref{feasibility_problem} augments the OPF constraints \eqref{eq:qc_p}--\eqref{eq:qc_others} with the separating hyperplane constraint such that the resulting disconnectedness certificates consider all OPF constraints.}


Infeasibility of~\eqref{feasibility_problem} guarantees that all points on the hyperplane are infeasible. Applying bound tightening techniques, as described in Section~\ref{subsec:bt}, identifies cases where the relaxed formulation~\eqref{feasibility_problem} is infeasible, thus certifying the disconnectedness for the corresponding OPF feasible spaces. Conversely, if applying bound tightening to~\eqref{feasibility_problem} does not prove infeasibility, then the test for disconnectedness is indeterminate. Rotating the hyperplane in~\eqref{feasibility_problem} may help certify infeasibility in such cases, as shown numerically in Section~\ref{Numerical_results} via ad hoc rotations.

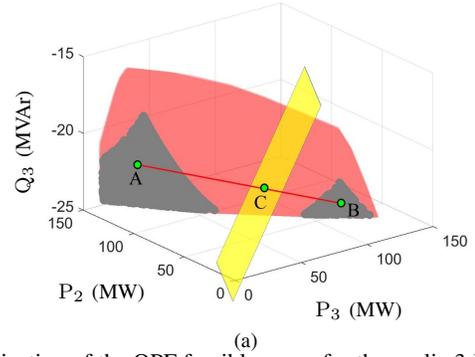
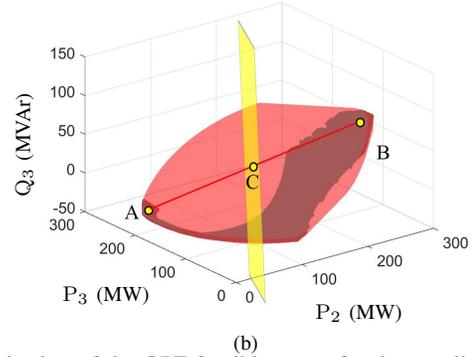
\begin{figure}[t]
\centering

\subfloat[]{%
  \begin{overpic}[width=6cm,height=5.6cm,keepaspectratio,tics=10]{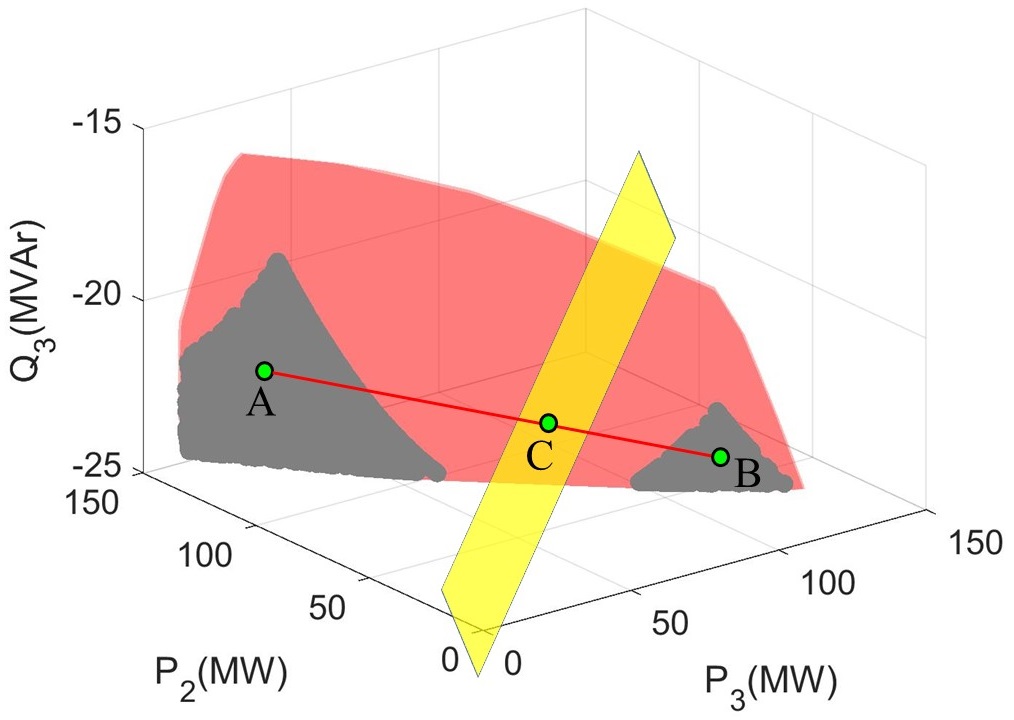}
    \put(10,-1){\color{white}\rule{4em}{1.5em}}
    \put(-2,30){\color{white}\rule{1.5em}{4em}}
    \put(65,-2){\color{white}\rule{4em}{1.5em}}
    \put(18,6){\makebox(0,0){\footnotesize  $\mathrm{P_2}$ (MW)}}
    \put(-1,28){\rotatebox{90}{\footnotesize  $\mathrm{Q_3}$ (MVAr)}}
    \put(75,2){\makebox(0,0){\footnotesize  $\mathrm{P_3}$ (MW)}}
  \end{overpic}
  \label{f:FS_cyclicthreebusFS}
}

\parbox{\columnwidth}{\small
(a) A projection of the OPF feasible space for the cyclic 3-bus system in~\cite{Narimani_ACC}.  
The proposed algorithm certifies the space's disconnectedness.
}


\subfloat[]{%
  \begin{overpic}[width=6cm,height=5.6cm,keepaspectratio,tics=10]{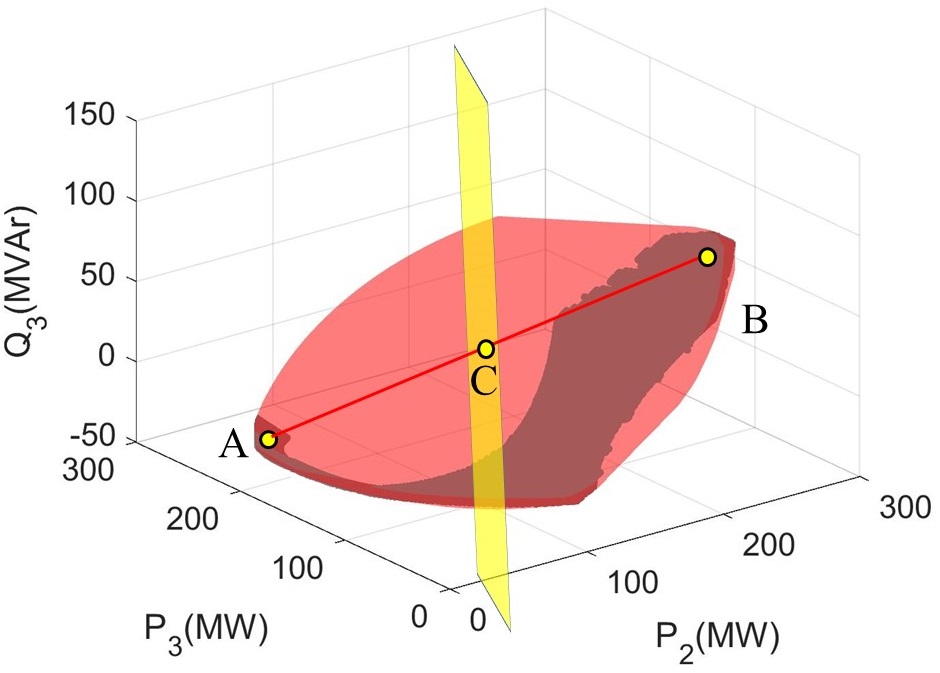}
    \put(10,-1){\color{white}\rule{4em}{1.5em}}
    \put(-3,30){\color{white}\rule{1.5em}{4em}}
    \put(65,-2){\color{white}\rule{4em}{1.5em}}
    \put(19,6){\makebox(0,0){\footnotesize  $\mathrm{P_3}$ (MW)}}
    \put(-1,28){\rotatebox{90}{\footnotesize  $\mathrm{Q_3}$ (MVAr)}}
    \put(75,2){\makebox(0,0){\footnotesize  $\mathrm{P_2}$ (MW)}}
  \end{overpic}
  \label{f:FS_AcyclicthreebusFS}
}

\parbox{\columnwidth}{\small
(b) A projection of the OPF feasible space for the acyclic 3-bus test case in~\cite{Narimani_ACC}.  
The OPF feasible space is non-convex but connected, and the proposed algorithm does not certify disconnectedness.
}


\caption{Illustrative examples of applying the proposed algorithm to two test cases from~\cite{Narimani_ACC}. 
The OPF feasible spaces, which are computed using the method in~\cite{molzahn_opf_spaces}, are shown in gray. The pink regions correspond to the QC relaxation's feasible spaces (without bound tightening). The method from~\cite{molzahn-nonconvexity_search} identifies non-convexities in both feasible spaces via the feasible points $A$ and $B$ and the infeasible point $C$ on the line connecting $A$ and $B$. The yellow hyperplane is normal to the line connecting $A$ and $B$ and passes through point $C$. Applying bound tightening to~\eqref{feasibility_problem} certifies infeasibility of all points on the yellow hyperplane in Fig.~\ref{f:FS_cyclicthreebusFS}, and thus this feasible space is disconnected. Conversely, the OPF problem's feasible space intersects the yellow hyperplane in Fig.~\ref{f:FS_AcyclicthreebusFS} and thus~\eqref{feasibility_problem} is feasible. Accordingly, the proposed algorithm does not indicate disconnectedness of this OPF feasible space.}
\label{fig:threebusFS}
\vspace*{-1.75em}
\end{figure}

To illustrate the proposed algorithm, Fig.~\ref{fig:threebusFS} visualizes results from two three-bus test cases from~\cite{Narimani_ACC}. In Fig.~\ref{f:FS_cyclicthreebusFS}, the proposed algorithm certifies that the OPF feasible space is disconnected via infeasibility of~\eqref{feasibility_problem} for the yellow hyperplane, after bound tightening. Conversely, the OPF feasible space in Fig.~\ref{f:FS_AcyclicthreebusFS} is non-convex but connected. Accordingly, the algorithm does not certify disconnectedness since the yellow plane passes through the OPF feasible space, and thus~\eqref{feasibility_problem} is feasible.

\begin{figure}[t]
  \centering
  \begin{overpic}[width=0.4\textwidth,tics=10]{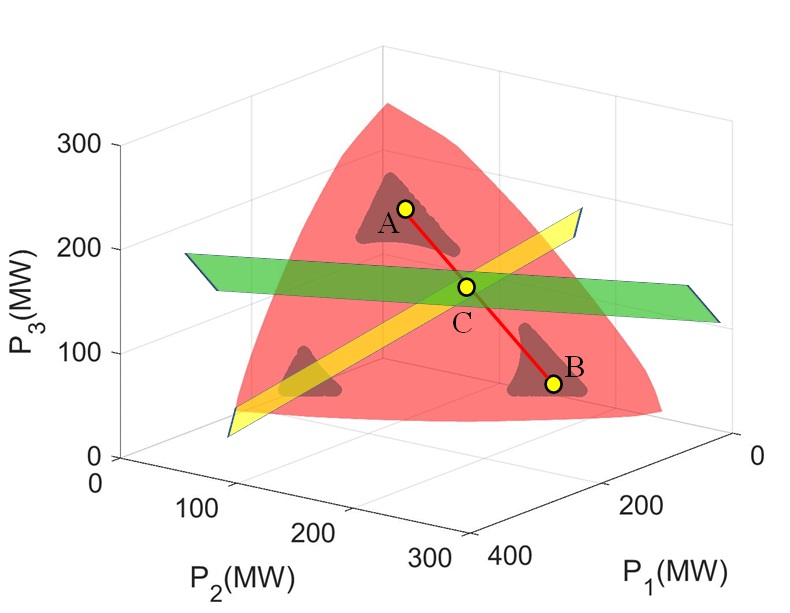}
    \put(18,1){\color{white}\rule{4em}{1.7em}}   
    \put(-3,30){\color{white}\rule{2em}{4em}}   
    \put(76,0){\color{white}\rule{4em}{1.5em}}  

    \put(85,6){\makebox(0,0){\small  $\mathrm{P_1~(MW)}$}}
    \put(0,30){\rotatebox{90}{\small  $\mathrm{P_3~(MW)}$}}
    \put(32,4){\makebox(0,0){\small  $\mathrm{P_2~(MW)}$}}
  \end{overpic}
  \caption{Feasible space projection for the 9-bus test system from~\cite{bukhsh_tps}. Although the OPF problem's feasible space shown by the gray regions is disconnected, the proposed algorithm fails to certify disconnectedness since the yellow plane that is normal to the line segment between point $A$ and $B$ passes through the component of the feasible space in the lower-left corner of the figure. Conversely, applying bound tightening to the green hyperplane which is rotated by $45^\circ$ around the $P_2$ axis certifies disconnectedness.}
  \label{fig:FS_9bus}
  \vspace{-1.5em}
\end{figure}
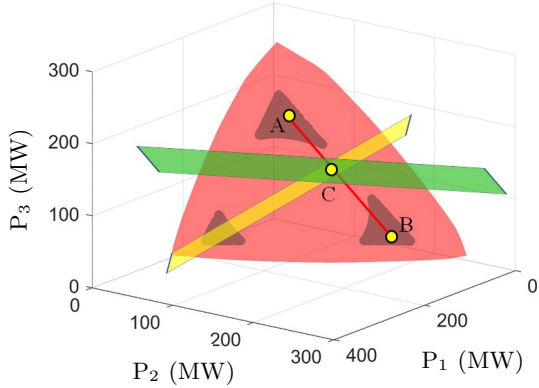

Fig.~\ref{fig:FS_9bus} shows another example using the nine-bus system from~\cite{bukhsh_tps} which has a feasible space with three disconnected components. The method from~\cite{molzahn-nonconvexity_search} identifies points $A$ and $B$ as feasible with the infeasible point $C$ on the connecting line segment. In this example, the normal hyperplane at point $C$, shown in yellow, does not certify disconnectedness as it passes through the third disconnected component of the feasible space in the lower-left corner of the figure. Conversely, applying bound tightening using the green hyperplane, which rotates the yellow hyperplane by $45^\circ$ around the $P_2$ axis, does certify disconnectedness of the feasible space. 

\rn{This particular rotation was identified via trial and error. Future work will focus on developing a margin-maximizing orientation strategy to automate the selection of the separating hyperplane. Instead of using manual rotations, as in Fig.~\ref{fig:FS_9bus}, this approach will treat the hyperplane’s orientation as an optimization variable and search for the direction that maximizes an infeasibility measure, such as the minimum aggregate violation of the OPF constraints on the QC-relaxed feasibility problem. A hierarchical coarse-to-fine search, combining a broad grid scan with local derivative-free refinement, could help find the orientation that provides the largest infeasibility margin, thereby improving the algorithm’s robustness.}

\begin{table*}[t]
\rn{
\centering
\caption{\rn{Comparison Between Feasible Path Algorithms and the Proposed Disconnectedness-Certifying Algorithm}}
\centering
\footnotesize
\renewcommand{\arraystretch}{0.95}
\setlength{\tabcolsep}{5pt}
\label{table:comparison}
\begin{tabular}{
    m{0.54cm}
    >{\centering\arraybackslash}m{2.1cm}
    >{\centering\arraybackslash}m{1.65cm}
    >{\centering\arraybackslash}m{1.95cm}
    >{\hspace{-1em}}m{10cm}
}
\toprule
\textbf{Ref.} &
\textbf{Certifies Disconnectedness} &
\textbf{Identifies a Feasible Path} &
\textbf{Optimizes Control Actions} &
\multicolumn{1}{c}{\textbf{Core Principle}} \\[-0.25em]
\midrule
\textbf{\cite{lee2020}} & 
\xmark & \cmark & \xmark &
Uses sequential convex restrictions to trace feasible paths between operating points. \\
\hline
\textbf{\cite{barros2022}} & 
\xmark & \cmark & \cmark &
Sequentially adjusts controls using Lagrange multiplier sensitivities to minimize losses. \\
\hline
\textbf{\cite{turizo2025}} & 
\xmark & \cmark & \cmark &
Finds a discretized shortest path via a nonlinear program that minimizes control actuations. \\
\hline
\textbf{Proposed\newline Algorithm\hspace{-2em}} & 
\cmark & \xmark & \xmark &
Certifies disconnectedness by proving infeasibility of all points on a separating hyperplane. \\
\bottomrule
\end{tabular}}
\vspace{-1em}
\end{table*}


\rn{The comparison in Table~\ref{table:comparison} emphasizes the fundamental differences between existing feasible path algorithms and the proposed disconnectedness-certifying approach.
While prior literature~\cite{lee2020,barros2022,turizo2025} focus on constructing feasible transition paths between operating points, none can rigorously determine when such a path does not exist.
To the best of our knowledge, our proposed algorithm is the first and only algorithm capable of certifying the disconnectedness of an OPF feasible space.
Thus, the proposed approach complements existing path-finding methods by definitely certifying when feasible transitions between operating points are impossible.}

\section{Numerical Results}
\label{Numerical_results}

\rn{This section summarizes numerical results from applying the proposed algorithm to challenging OPF test cases from~\cite{pglib, Narimani_ACC, bukhsh_tps}. These numerical studies were performed on a computer with a 3.3 GHz Intel Core i7 processor and 16 GB of RAM. The proposed algorithm is implemented in \mbox{MATLAB} using \mbox{YALMIP} (R2023.09)~\cite{yalmip} and solved using MOSEK (10.0.46). Default interior-point settings were employed, with primal/dual feasibility and relative/absolute optimality gap tolerances of $10^{-6}$. For the optimization-based bound-tightening procedure, the algorithm was executed iteratively until no variable bound improved beyond $10^{-4}$~p.u. for power injections and $10^{-5}$~p.u. for voltage magnitudes.}

\begin{table}[t!]
\caption{Results for checking the feasibility of transitioning between different points in various test cases.}
    \centering
    \begin{tabular}{|p{2.09cm}|c|c|}
\hline 
\thead{Test\\ Cases}  & \thead{Rotation\\ Angles} & \thead{Disconnected\\ Feasible Space}\tabularnewline
\hline 
\hline 
Cyclic 3-bus~\cite{Narimani_ACC}     &  $(0^\circ,0^\circ,0^\circ)$& Yes \tabularnewline
\hline 
5-bus~\cite{Narimani_ACC}     &  $(0^\circ,0^\circ,0^\circ)$& Yes \tabularnewline
\hline 
5-bus~\cite{bukhsh_tps}     &  $(0^\circ,0^\circ,0^\circ)$& Yes \tabularnewline
\hline 
 9-bus~\cite{bukhsh_tps}     &  $(0^\circ,45^\circ,0^\circ)$&  Yes\tabularnewline
 \hline 
 14-bus-sad~\cite{pglib}     &  $(0^\circ,0^\circ,0^\circ)$& Yes\tabularnewline
\hline 
 30-bus-ieee~\cite{pglib}     &  $(0^\circ,0^\circ,0^\circ)$& Yes\tabularnewline
\hline 
 118-bus-api~\cite{bukhsh_tps}     &  $(0^\circ,0^\circ,0^\circ)$& Yes\tabularnewline
\hline 
\end{tabular}
    \label{tab:Results}
\vspace{-1em}
\end{table}

Table~\ref{tab:Results} shows selected results obtained by applying the proposed algorithm to different test cases. The proposed algorithm identifies disconnectedness for several test cases, ranging from very small to mid-size systems. For the small cases, visualizing the feasible spaces as in~\cite{Narimani_ACC} (see Figs.~\ref{f:FS_cyclicthreebusFS} and~\ref{fig:FS_9bus} for two examples) corroborates the disconnectedness certificates from our proposed algorithm. For the test cases in Table~\ref{tab:Results}, the voltage phasors for the feasible points which belong to different disconnected components of the feasible space are given in~\ifarxiv the appendix\else \cite[Appendix]{narimani2025certifyingnonexistencefeasiblepath}\fi. These test cases are known to challenge OPF solution algorithms, with many relaxations of these problems yielding large relaxation gaps and/or local solvers finding suboptimal local solutions. This is consistent with our expectation that OPF algorithms often struggle with problems that have disconnected feasible spaces.



For other test cases, the proposed algorithm returned an indeterminate result, i.e., did not certify disconnectedness of the OPF feasible spaces. We note that this does not necessarily indicate that the feasible spaces for these problems are connected. Different combinations of points $A$, $B$, and $C$ and/or different orientations of the candidate separating hyperplanes may potentially be able to certify disconnectedness, as our condition is sufficient but not necessary.

\rn{Our algorithm does not certify disconnectedness for many PGLib test cases~\cite{lee2020}, which suggests but does not guarantee that their feasible spaces may be connected. We know from the feasible path algorithm in~\cite{lee2020} that feasible paths exist between certain pairs of operating points considered in that paper for systems such as ``57-bus-ieee'' and ``118-bus''. However, our indeterminate result means that it is unclear whether the feasible spaces for these problems are indeed connected everywhere or if the pairs of operating points in~\cite{lee2020} happen to be in the same component of a disconnected feasible space.}%

Lastly, we discuss the computational aspects of the proposed algorithm. The most computationally demanding step is the application of optimization-based bound tightening to certify infeasibility of~\eqref{feasibility_problem}. However, for all the OPF problems we tested, the bound tightening step was not computationally intensive relative to typical bound tightening applications such as global solution methods for OPF problems~\cite{coffrin2016strengthen_tps,StrongSOCPRelaxations,chen2015,arctan2}. \rn{In our algorithm, the bound-tightening step typically requires only a few iterations to either certify the infeasibility of transitioning between the candidate operating points or terminate without certification.
For example, the algorithm certifies the disconnectedness for the ``118-bus-api'' test case in approximately six minutes, whereas traditional optimization-based bound-tightening methods applied in global OPF solvers often require substantially longer runtimes, on the order of hours, for systems of comparable size.
Recent bound-tightening advances have further improved scalability, making such techniques tractable even for large OPF problems~\cite{shchetinin2019efficient,cengil2025learning}.
Moreover, unlike conventional bound tightening settings, the proposed algorithm can certify transition infeasibility within only a few iterations, maintaining computational efficiency for relatively large systems. We attribute this to the fact that the feasibility problem~\eqref{feasibility_problem} restricts the QC relaxation to a hyperplane rather than the entire OPF feasible space. Future work will incorporate more scalable bound-tightening and parallel computing approaches~\cite{shchetinin2019efficient,cengil2025learning} to further improve tractability.}%


\section{Conclusion}
\label{conclusion}

\rn{This paper presents an algorithm for certifying the presence of disconnected components in the feasible spaces of OPF problems. The algorithm identifies candidate operating points that may lie in different components and employs the QC relaxation with bound tightening to verify that all points on a separating hyperplane are infeasible. When infeasibility is confirmed, the two operating points are proven to belong to distinct disconnected components, indicating that no feasible path exists between them. 
To the best of our knowledge, this is the first algorithm capable of rigorously certifying disconnectedness in OPF feasible spaces. Understanding whether non-convexities correspond to disconnected components enhances both theoretical insights into OPF feasible space geometry and the reliability of power system operations, with practical implications by enabling feasible-space decomposition, contingency-aware dispatch, and secure operating point selection. 
Future work will focus on developing systematic methods for selecting the optimal rotation of the separating hyperplane and integrating this approach with feasible path algorithms.}

%

\ifarxiv
\appendix
\label{apx:Feasible_Points_Voltages}
Tables~II--VIII provide the voltage phasors for points $A$ and $B$ which our proposed algorithm identifies as belonging to different disconnected components of the corresponding OPF problems' feasible spaces.

\begin{table}
\centering
\caption{Voltages for points $A$ and $B$ for the 3-bus test system in~\cite{Narimani_ACC}.}
    \begin{tabular}{|c|c|c|}
\hline 
 & Point A & Point B\tabularnewline
\hline 
\hline 
1 & 0.8962 - 0.0919i & 0.8846 - 0.1660i\tabularnewline
\hline 
2 & 0.9040 + 0.0000i & 0.9996 + 0.0000i\tabularnewline
\hline 
3 & 1.0282 + 0.0888i & 0.9278 - 0.0740i\tabularnewline
\hline 
\end{tabular}
\end{table}

\begin{table}
\centering
\caption{Corresponding Voltages for points $A$ and $B$ for the 5-bus test system in~\cite{Narimani_ACC}.}
    \begin{tabular}{|c|c|c|}
\hline 
Bus & Point A & Point B\tabularnewline
\hline 
\hline 
1 & 0.9184 - 0.1354i & 0.8463 - 0.3785i\tabularnewline
\hline 
2 & 0.9030 - 0.1643i & 0.8436 - 0.4049i\tabularnewline
\hline 
3 & 0.9878 + 0.0000i & 1.0878 + 0.0000i\tabularnewline
\hline 
4 & 0.9040 - 0.1524i & 0.8370 - 0.3940i\tabularnewline
\hline 
5 & 1.0495 + 0.2700i & 0.9428 - 0.2943i\tabularnewline
\hline 
\end{tabular}
\end{table}

\begin{table}
    \centering
    \caption{Voltages for points $A$ and $B$ for the 5-bus test system in~\cite{bukhsh_tps}.}
    \begin{tabular}{|c|c|c|}
\hline 
 & Point A & Point B\tabularnewline
\hline 
\hline 
1 & 1.0467 - 0.0000i & 1.0500 - 0.0000i\tabularnewline
\hline 
2 & 0.9550 - 0.0578i & 0.9838 - 0.1183i\tabularnewline
\hline 
3 & 0.9485 - 0.0533i & 0.9796 - 0.1196i\tabularnewline
\hline 
4 & 0.7791 + 0.6011i & 0.9931 - 0.1650i\tabularnewline
\hline 
5 & 0.7362 + 0.7487i & 1.0317 - 0.1032i\tabularnewline
\hline 
\end{tabular}
\end{table}

\begin{table}
    \centering
    \caption{Voltages for points $A$ and $B$ for the 9-bus test system in~\cite{bukhsh_tps}.}
\begin{tabular}{|c|c|c|}
\hline 
 & Point A & Point B\tabularnewline
\hline 
\hline 
1 & 0.9025 + 0.0000i & 0.9058 + 0.0000i\tabularnewline
\hline 
2 & 0.9042 - 0.1541i & 0.9220 + 0.0789i\tabularnewline
\hline 
3 & 0.9063 - 0.1819i & 0.9082 + 0.2334i\tabularnewline
\hline 
4 & 0.9052 - 0.0927i & 0.9089 - 0.0063i\tabularnewline
\hline 
5 & 0.8961 - 0.1606i & 0.9103 + 0.0105i\tabularnewline
\hline 
6 & 0.9082 - 0.1888i & 0.9297 + 0.1444i\tabularnewline
\hline 
7 & 0.8934 - 0.2080i & 0.9246 + 0.0647i\tabularnewline
\hline 
8 & 0.9034 - 0.1787i & 0.9275 + 0.0545i\tabularnewline
\hline 
9 & 0.8848 - 0.1644i & 0.8995 - 0.0302i\tabularnewline
\hline 
\end{tabular}
\end{table}

\begin{table}
\caption{Voltages for points $A$ and $B$ for pglib 14-bus-ieee-sad.}
    \centering
\begin{tabular}{|c|c|c|}
\hline 
Bus & Point A & Point B \tabularnewline
\hline 
\hline 
1 & 1.0600 + 1.1758i & 1.0600 + 0i \tabularnewline
\hline 
2 & 1.0348 - 0.0785i & 1.0351 - 0.0785i\tabularnewline
\hline 
3 & 0.9930 - 0.2185i & 0.9745 - 0.2108i\tabularnewline
\hline 
4 & 0.9967 - 0.1730i & 0.9923 - 0.1718i\tabularnewline
\hline 
5 & 1.0028 - 0.1470i & 1.0001 - 0.1465i\tabularnewline
\hline 
6 & 1.0290 - 0.2546i & 1.0289 - 0.2551i\tabularnewline
\hline 
7 & 1.0180 - 0.2339i & 1.0160 - 0.2333i\tabularnewline
\hline 
8 & 1.0331 - 0.2373i & 1.0331 - 0.2372i\tabularnewline
\hline 
9 & 1.0080 - 0.2617i & 1.0061 - 0.2611i\tabularnewline
\hline 
10 & 1.0031 - 0.2637i & 1.0015 - 0.2633i\tabularnewline
\hline 
11 & 1.0120 - 0.2607i & 1.0110 - 0.2606i\tabularnewline
\hline 
12 & 1.0102 - 0.2663i & 1.0100 - 0.2667i\tabularnewline
\hline 
13 & 1.0049 - 0.2663i & 1.0045 - 0.2665i\tabularnewline
\hline 
14 & 0.9842 - 0.2767i & 0.9830 - 0.2765i\tabularnewline
\hline 
\end{tabular}
\end{table}

\begin{table}
\caption{Voltages for points $A$ and $B$ for pglib 30-bus-ieee.}
\centering
\begin{tabular}{|c|c|c|}
\hline 
Bus & Point A & Point B \tabularnewline
\hline 
\hline 
1 & 1.0600 +0i & 1.0600 + 0i\tabularnewline
\hline 
2 & 1.0405 - 0.0707i & 1.0350 - 0.0687i\tabularnewline
\hline 
3 & 1.0077 - 0.1168i & 1.0081 - 0.1174i\tabularnewline
\hline 
4 & 0.9935 - 0.1422i & 0.9940 - 0.1429i\tabularnewline
\hline 
5 & 0.9816 - 0.2249i & 0.9805 - 0.2248i\tabularnewline
\hline 
6 & 0.9841 - 0.1698i & 0.9875 - 0.1721i\tabularnewline
\hline 
7 & 0.9739 - 0.2000i & 0.9755 - 0.2013i\tabularnewline
\hline 
8 & 0.9760 - 0.1803i & 0.9857 - 0.1856i\tabularnewline
\hline 
9 & 1.0088 - 0.2315i & 1.0088 - 0.2334i\tabularnewline
\hline 
10 & 0.9973 - 0.2591i & 0.9957 - 0.2608i\tabularnewline
\hline 
11 & 1.0332 - 0.2371i & 1.0327 - 0.2390i\tabularnewline
\hline 
12 & 1.0149 - 0.2510i & 1.0074 - 0.2492i\tabularnewline
\hline 
13 & 1.0290 - 0.2545i & 1.0133 - 0.2507i\tabularnewline
\hline 
14 & 0.9959 - 0.2630i & 0.9891 - 0.2616i\tabularnewline
\hline 
15 & 0.9907 - 0.2631i & 0.9847 - 0.2623i\tabularnewline
\hline 
16 & 0.9988 - 0.2572i & 0.9937 - 0.2571i\tabularnewline
\hline
17 & 0.9918 - 0.2611i & 0.9891 - 0.2622i\tabularnewline
\hline 
18 & 0.9778 - 0.2707i & 0.9733 - 0.2709i\tabularnewline
\hline 
19 & 0.9742 - 0.2727i & 0.9705 - 0.2734i\tabularnewline
\hline 
20 & 0.9789 - 0.2702i & 0.9758 - 0.2711i\tabularnewline
\hline 
21 & 0.9832 - 0.2638i & 0.9814 - 0.2655i\tabularnewline
\hline 
22 & 0.9838 - 0.2638i & 0.9820 - 0.2653i\tabularnewline
\hline
23 & 0.9781 - 0.2666i & 0.9737 - 0.2667i\tabularnewline
\hline 
24 & 0.9710 - 0.2674i & 0.9690 - 0.2686i\tabularnewline
\hline 
25 & 0.9692 - 0.2586i & 0.9695 - 0.2610i\tabularnewline
\hline 
26 & 0.9499 - 0.2612i & 0.9503 - 0.2635i\tabularnewline
\hline 
27 & 0.9775 - 0.2509i & 0.9794 - 0.2538i\tabularnewline
\hline 
28 & 0.9773 - 0.1796i & 0.9818 - 0.1824i\tabularnewline
\hline 
29 & 0.9524 - 0.2670i & 0.9543 - 0.2698i\tabularnewline
\hline 
30 & 0.9368 - 0.2787i & 0.9388 - 0.2815i\tabularnewline
\hline 
\end{tabular}    
\end{table}

\begin{table*}
\caption{Voltages for points $A$ and $B$ for pglib 118-bus-api.}
\scriptsize
\centering
\begin{tabular}{|c|c|c||c|c|c|}
\hline 
Bus & Point A & Point B & Bus & Point A & Point B\tabularnewline
\hline 
\hline 
1 & 1.0197 + 0.2895i & 1.0155 + 0.3040i & 60 & 0.9462 + 0.1948i & 0.9256 + 0.1967i\tabularnewline
\hline 
2 & 0.9970 + 0.3292i & 0.9849 + 0.3437i & 61 & 0.9459 + 0.2144i & 0.9257 + 0.2162i\tabularnewline
\hline 
3 & 1.0049 + 0.3088i & 0.9985 + 0.3230i & 62 & 0.9480 + 0.1671i & 0.9317 + 0.1728i\tabularnewline
\hline 
4 & 0.9918 + 0.3741i & 0.9865 + 0.3879i & 63 & 0.9396 + 0.2273i & 0.9199 + 0.2274i\tabularnewline
\hline 
5 & 0.9710 + 0.3807i & 0.9745 + 0.3942i & 64 & 0.9559 + 0.2148i & 0.9457 + 0.2154i\tabularnewline
\hline 
6 & 1.0001 + 0.3511i & 0.9948 + 0.3660i & 65 & 1.0004 + 0.1966i & 1.0157 + 0.1954i\tabularnewline
\hline 
7 & 0.9952 + 0.3558i & 0.9854 + 0.3705i & 66 & 1.0015 + 0.1488i & 1.0038 + 0.1759i\tabularnewline
\hline 
8 & 0.9065 + 0.4304i & 0.9036 + 0.4427i & 67 & 0.9720 + 0.1310i & 0.9661 + 0.1480i\tabularnewline
\hline 
9 & 0.8068 + 0.6162i & 0.8014 + 0.6266i & 68 & 0.9952 + 0.0880i & 1.0025 + 0.0886i\tabularnewline
\hline 
10 & 0.6687 + 0.7912i & 0.6608 + 0.7992i & 69 & 0.9918 + 0i & 0.9932 + 0i\tabularnewline
\hline 
11 & 0.9820 + 0.3455i & 0.9698 + 0.3594i & 70 & 1.0461 - 0.1712i & 1.0462 - 0.1706i\tabularnewline
\hline 
12 & 0.9867 + 0.3744i & 0.9694 + 0.3890i & 71 & 1.0484 - 0.1566i & 1.0485 - 0.1560i\tabularnewline
\hline 
13 & 0.9665 + 0.2559i & 0.9571 + 0.2686i & 72 & 1.0577 - 0.0705i & 1.0577 - 0.0610i\tabularnewline
\hline 
14 & 0.9784 + 0.2891i & 0.9655 + 0.3024i & 73 & 1.0476 - 0.1591i & 1.0477 - 0.1585i\tabularnewline
\hline 
15 & 0.9506 + 0.1009i & 0.9491 + 0.1109i & 74 & 0.9190 - 0.1975i & 0.9191 - 0.1971i\tabularnewline
\hline 
16 & 0.9756 + 0.2856i & 0.9635 + 0.2985i & 75 & 0.9404 - 0.1802i & 0.9398 - 0.1795i\tabularnewline
\hline 
17 & 0.9610 + 0.1708i & 0.9590 + 0.1809i & 76 & 0.9118 - 0.2284i & 0.9118 - 0.2284i\tabularnewline
\hline 
18 & 0.9824 + 0.0856i & 0.9746 + 0.0963i & 77 & 1.0029 - 0.1090i & 0.9940 - 0.1040i\tabularnewline
\hline 
19 & 0.9805 + 0.0584i & 0.9789 + 0.0682i & 78 & 0.9954 - 0.1175i & 0.9863 - 0.1126i\tabularnewline
\hline 
20 & 0.9744 + 0.0282i & 0.9701 + 0.0363i & 79 & 0.9977 - 0.1091i & 0.9885 - 0.1041i\tabularnewline
\hline 
21 & 0.9753 + 0.0326i & 0.9689 + 0.0397i & 80 & 1.0267 - 0.0427i & 1.0174 - 0.0369i\tabularnewline
\hline 
22 & 0.9865 + 0.0600i & 0.9774 + 0.0661i & 81 & 0.9840 + 0.0394i & 0.9852 + 0.0422i\tabularnewline
\hline 
23 & 1.0118 + 0.1318i & 0.9986 + 0.1366i & 82 & 0.9893 - 0.0722i & 0.9813 - 0.0666i\tabularnewline
\hline 
24 & 1.0254 + 0.0610i & 1.0174 + 0.0640i & 83 & 1.0056 - 0.0405i & 0.9998 - 0.0346i\tabularnewline
\hline 
25 & 1.0008 + 0.2774i & 0.9841 + 0.2838i & 84 & 1.0353 + 0.0245i & 1.0332 + 0.0309i\tabularnewline
\hline 
26 & 0.8751 + 0.3432i & 0.9534 + 0.3781i & 85 & 1.0580 + 0.0662i & 1.0575 + 0.0729i\tabularnewline
\hline 
27 & 1.0509 + 0.1388i & 1.0505 + 0.1414i & 86 & 1.0261 + 0.1930i & 1.0249 + 0.1995i\tabularnewline
\hline 
28 & 1.0130 + 0.1834i & 0.9950 + 0.1870i & 87 & 0.9539 + 0.4622i & 0.9510 + 0.4681i\tabularnewline
\hline 
29 & 0.9784 + 0.2605i & 0.9408 + 0.2660i & 88 & 1.0028 + 0.1004i & 0.9916 + 0.1087i\tabularnewline
\hline 
30 & 0.9333 + 0.2433i & 0.9453 + 0.2582i & 89 & 0.9751 + 0.1814i & 0.9561 + 0.1915i\tabularnewline
\hline 
31 & 0.9695 + 0.3026i & 0.9249 + 0.3099i & 90 & 0.9671 + 0.0070i & 0.9484 + 0.0141i\tabularnewline
\hline 
32 & 0.9892 + 0.1511i & 0.9655 + 0.1561i & 91 & 0.9550 + 0.0357i & 0.9390 + 0.0425i\tabularnewline
\hline 
33 & 0.9871 + 0.0101i & 0.9876 + 0.0205i & 92 & 0.9688 + 0.0850i & 0.9517 + 0.0934i\tabularnewline
\hline 
34 & 1.0587 - 0.0525i & 1.0592 - 0.0408i & 93 & 0.9613 + 0.0417i & 0.9462 + 0.0491i\tabularnewline
\hline 
35 & 1.0552 - 0.0668i & 1.0562 - 0.0552i & 94 & 0.9627 + 0.0187i & 0.9495 + 0.0256i\tabularnewline
\hline 
36 & 1.0579 - 0.0674i & 1.0585 - 0.0557i & 95 & 0.9544 - 0.0261i & 0.9423 - 0.0199i\tabularnewline
\hline 
37 & 1.0537 - 0.0393i & 1.0558 - 0.0281i & 96 & 0.9742 - 0.0484i & 0.9636 - 0.0425i\tabularnewline
\hline 
38 & 0.9781 + 0.0907i & 0.9855 + 0.1009i & 97 & 0.9939 - 0.0562i & 0.9839 - 0.0505i\tabularnewline
\hline 
39 & 1.0074 - 0.1595i & 1.0116 - 0.1490i & 98 & 0.9965 - 0.0230i & 0.9856 - 0.0168i\tabularnewline
\hline 
40 & 0.9980 - 0.2053i & 1.0031 - 0.1951i & 99 & 1.0412 + 0.0328i & 0.9385 + 0.0537i\tabularnewline
\hline 
41 & 0.9641 - 0.2171i & 0.9686 - 0.2068i & 100 & 0.9789 + 0.1096i & 0.9654 + 0.1178i\tabularnewline
\hline 
42 & 0.9230 - 0.1778i & 0.9250 - 0.1670i & 101 & 0.9609 + 0.0704i & 0.9458 + 0.0781i\tabularnewline
\hline 
43 & 1.0220 - 0.0402i & 1.0221 - 0.0287i & 102 & 0.9654 + 0.0769i & 0.9489 + 0.0850i\tabularnewline
\hline 
44 & 0.9930 + 0.0482i & 0.9917 + 0.0597i & 103 & 0.9633 + 0.0269i & 0.9292 + 0.1419i\tabularnewline
\hline 
45 & 0.9817 + 0.1064i & 0.9797 + 0.1179i & 104 & 0.9818 - 0.1019i & 0.9626 - 0.0061i\tabularnewline
\hline 
46 & 1.0142 + 0.3083i & 1.0105 + 0.3200i & 105 & 0.9713 - 0.1393i & 0.9518 - 0.0323i\tabularnewline
\hline 
47 & 0.9830 + 0.1464i & 0.9804 + 0.1567i & 106 & 0.9560 - 0.1421i & 0.9603 - 0.0606i\tabularnewline
\hline 
48 & 0.9764 + 0.1470i & 0.9732 + 0.1589i & 107 & 0.9584 - 0.2309i & 1.0458 - 0.1728i\tabularnewline
\hline 
49 & 0.9648 + 0.1224i & 0.9615 + 0.1343i & 108 & 0.9504 - 0.1871i & 0.9554 - 0.0420i\tabularnewline
\hline 
50 & 0.9564 + 0.1053i & 0.9541 + 0.1123i & 109 & 0.9424 - 0.2058i & 0.9572 - 0.0450i\tabularnewline
\hline 
51 & 0.9418 + 0.0853i & 0.9411 + 0.0862i & 110 & 0.9286 - 0.2463i & 0.9678 - 0.0429i\tabularnewline
\hline 
52 & 0.9371 + 0.0738i & 0.9372 + 0.0730i & 111 & 0.9286 - 0.2463i & 0.9384 + 0.0553i\tabularnewline
\hline 
53 & 0.9406 + 0.0943i & 0.9431 + 0.0889i & 112 & 0.8845 - 0.3181i & 0.9741 - 0.1373i\tabularnewline
\hline 
54 & 0.9587 + 0.1589i & 0.9634 + 0.1500i & 113 & 0.9243 + 0.1710i & 0.9227 + 0.1797i\tabularnewline
\hline 
55 & 0.9562 + 0.1386i & 0.9552 + 0.1313i & 114 & 1.0111 + 0.1300i & 0.9972 + 0.1338i\tabularnewline
\hline 
56 & 0.9561 + 0.1433i & 0.9566 + 0.1358i & 115 & 1.0151 + 0.1278i & 1.0030 + 0.1314i\tabularnewline
\hline 
57 & 0.9533 + 0.1152i & 0.9526 + 0.1138i & 116 & 1.0024 + 0.0805i & 1.0096 + 0.0812i\tabularnewline
\hline 
58 & 0.9453 + 0.1012i & 0.9450 + 0.0985i & 117 & 0.9825 + 0.3273i & 0.9660 + 0.3413i\tabularnewline
\hline 
59 & 0.9430 + 0.2561i & 0.9051 + 0.2539i & 118 & 0.9158 - 0.2145i & 0.9155 - 0.2141i\tabularnewline
\hline 
\end{tabular}
\end{table*}

\fi

\bibliographystyle{IEEEtran}
\ifarxiv
\IEEEtriggeratref{10}
\else
\IEEEtriggeratref{50}
\fi
\bibliography{ref}

\end{document}

\bibliographystyle{IEEEtran}
\bibliography{IEEEabrv,ref}

\end{document}